\documentclass[12pt]{svjour2}
\usepackage{amsmath,amssymb,ogonek}
\usepackage{graphicx,pstricks,pst-plot}
\include{epsf}

\def\(({\left(}
\def\)){\right)}                       
\def\[[{\left[}
\def\]]{\right]}

\newcommand{\E}{\mathbb{E}}
\newcommand{\bP}{\mathbb{P}}
\newcommand{\cP}{\mathcal{P}}
\newcommand{\cN}{\mathcal{N}}
\newcommand{\ud}{\mathrm{d}}
\newcommand{\cf}{{\it cf.~}}

\newrgbcolor{darkred}{.5 0 0}
\newrgbcolor{lightblue}{.9 1 .95}
\newrgbcolor{darkgreen}{0 .5 0}
\newrgbcolor{cyan}{.1 .5 1}
\newrgbcolor{watergreen}{.7 1 .8}
\newrgbcolor{waterred}{1 .6 .5}
\newrgbcolor{watermagenta}{1 .5 1}
\newrgbcolor{purple}{.6 .3 1}
\newrgbcolor{waterpurple}{.8 .5 1}

\newcommand{\be}{\begin{equation}}
\newcommand{\ee}{\end{equation}}
\newcommand{\bea}{\begin{eqnarray}}
\newcommand{\eea}{\end{eqnarray}}

\newcommand{\bs}{ \mbox{\boldmath$\sigma$}}
\newcommand{\bpi}{ \mbox{\boldmath$\pi$}}

\begin{document}
\title{Random subcubes as a toy model for constraint satisfaction problems}

\author{Thierry Mora \and Lenka Zdeborov\'a}

\institute{T. Mora \at
Lewis Sigler Institute for Integrative Genomics, Princeton University, Princeton, NJ 08544, USA
\and
L. Zdeborov\'a \at
Universit\'e Paris-Sud, LPTMS, UMR8626,  B\^at. 100, Universit\'e
Paris-Sud 91405 Orsay cedex, France \\
CNRS, LPTMS, UMR8626, B\^at. 100, Universit\'e Paris-Sud 91405 Orsay cedex, France
}

\date{Received: date / Accepted: date}

\maketitle

\begin{abstract} 
We present an exactly solvable random-subcube model inspired by the structure of hard constraint satisfaction and optimization problems. Our model reproduces the structure of the solution space of the random $k$-satisfiability and $k$-coloring problems, and undergoes the same phase transitions as these problems.
The comparison becomes quantitative in the large-$k$ limit. Distance properties, as well the $x$-satisfiability threshold, are studied. The model is also generalized to define a continuous energy landscape useful for studying several aspects of glassy dynamics.

\keywords{Constraint satisfaction problems \and Clustering of solutions \and Exactly solvable models}
\end{abstract}
%

%

\section{Introduction}

Combinatorial optimization and Constraint Satisfaction Problems (CSPs) arise in a wide array of scientific branches, including statistical physics, information theory, inference and machine learning. These problems, which involve a large number of variables interacting through a large number of constraints or cost terms, are in general very hard to solve, and in most cases no algorithm seems to be able to find a solution within a reasonable time, as formalized by the $P\neq NP$ conjecture \cite{Papadimitriou94}.

In order to circumvent this intrinsic difficulty and possibly to identify (or avoid) ``hard'' instances, random ensembles of optimization problems were introduced and used as test beds for theories and algorithms. 
This line of research has considerably benefited from the methods and concepts of statistical mechanics \cite{KirkpatrickSelman94,MonassonZecchina99,BiroliMonasson00,MezardParisi02}.
In particular, a spectacular breakthrough was made by the development of the survey propagation algorithm \cite{MezardParisi02,MezardZecchina02} which is able to solve large random instances of CSPs in the so-called ``hard-SAT'' region.
The key to the success of the physics approach lies in the understanding of the rugged energy landscape (reminiscent of glassy phases) exhibited by these problems, which survey propagation exploits and integrates into a sophisticated message-passing \cite{KschischangFrey01} scheme.

The structure and organisation of solutions has been analyzed in detail for several CSPs, including the satisfiability problem ($k$-SAT) \cite{KrzakalaMontanari06}, the colorability of random graphs ($k$-COL) \cite{MezardPalassini05,KrzakalaMontanari06,ZdeborovaKrzakala07}, and systems of linear Boolean equations ($k$-XORSAT) \cite{CoccoDubois03,MezardRicci03,MoraMezard06}. It was shown that as the density of constraints is increased the space of solutions undergoes several phase transitions. At low density of constraints the solution space is concentrated in one big ergodic component, called ``cluster'' or ``state''.
For higher densities the systems undergoes a clustering transition, whereby the solution space breaks into an exponential number of well-separated clusters (this separation can be energetic or entropic). For even higher densities a second transition occurs, which reduces these clusters to a finite number. Finally, all clusters disappear at the SAT-UNSAT threshold. 

Since the clustering phenomenon is one of the main building blocks underlying the statistical physics approach\footnote{Technically, clustering is closely related to the one-step replica-symmetry breaking ansatz used in the replica/cavity method \cite{MezardParisi01,MezardZecchina02}.},
substantial efforts have been made to give a rigorous base to it \cite{MezardMora05,AchlioptasRicci06}. 
Some mathematical results were also obtained in the ergodic phase \cite{MontanariShah06}, and in the simple case of linear Boolean equations \cite{DuboisMandler02}.
Remarkably, pure states play an central role in spin-glass theory, and they have been extensively studied in that context.
However, the geometrical organization of glassy phases is not yet fully understood, and the classical picture of complex energy landscapes with many ``valleys'' still lacks an appropriate representation.



In this paper we introduce an exactly solvable random-subcube model\footnote{Independently of our work A. Montanari inspired by an idea of D. Achlioptas also introduced this model and worked out some parts of our Sec. \ref{one}.} (RSM), in the spirit of Derrida's Random Energy Model (REM) \cite{Derrida80}. This model
is inspired by the structure of hard CSPs and optimization problems, and reproduces most of their phenomenology. 
It can be also thought of as an attempt to construct a minimal setting that is able to reproduce the structure of solutions in hard CSPs.
Its purpose is mainly pedagogical, and it offers an excellent testing playground for ideas and methods in combinatorial optimization and glass physics, while being fully tractable.

Despite its simplicity, the RSM undergoes the same phase transitions as those observed in random CSPs such as $k$-SAT \cite{MezardZecchina02,KrzakalaMontanari06} or $k$-COL \cite{KrzakalaMontanari06,ZdeborovaKrzakala07}. The connection even becomes quantitative in the large-$k$ limit of these problems. So far only the zeroth order of this limit was intuitively known and related to Shannon's random code model \cite{Shannon48,Montanari01,BargForney02}, in which clusters are uniformly distributed singletons. One of our most notable results is that the RSM provides the first-order approximation of this large-$k$ limit, reproducing the cluster size distribution and freezing properties of the original models.

We also generalize the RSM to deal with continuous energy landscapes resembling those observed in glassy systems and hard optimization problems, and show how static and dynamical properties can be related explicitly. This energetic RSM displays temperature chaos, undergoes a dynamical transition, and has a Kauzmann temperature.

The paper is organized as follows: In Sec.~\ref{one} we define the model and describe its basic properties, as well as the connection with the random $k$-SAT and $k$-COL problems. We also analyze the behaviour of physics-guided decimation schemes. In Sec.~\ref{two} we discuss the relation between its dynamical and geometrical properties. In Sec.~\ref{three} we extend the definition to energetic landscapes, compute static and dynamical properties, and comment on some ideas from the physics of glassy systems. Finally we present a general discussion of our results in Sec.~\ref{conclusions}.    

\section{The random-subcube model}
\label{one}

\subsection{Definition}

Most constraint satisfaction problems are defined by a set of constraints on $N$ variables $\bs=(\sigma_1, \ldots, \sigma_N)$ with a finite alphabet, e.g. $\{0,1\}$. In contrast, the random-subcube model is defined directly by its solution space $S\subset\{0,1\}^N$; we define $S$ as the union of $\lfloor 2^{(1-\alpha)N}\rfloor$ random clusters (where $\lfloor x \rfloor$ denotes the integer value of $x$). A random cluster $A$ being defined as:
\be \label{eq:defcluster}
A=\{\vec \sigma\ |\ \forall i \in  \{1,\ldots,N\}, \sigma_i\in \pi^A_i\},
\ee
where $\bpi^A$ is a random mapping:
\begin{eqnarray}
\pi^A: \{1,\ldots,N\} & \longrightarrow & \{\{0\},\{1\},\{0,1\}\} \\
i & \longmapsto & \pi^A_i
\end{eqnarray}
such that for each variable $i$, $\pi^A_i=\{0\}$ with probability $p/2$, $\{1\}$ with probability $p/2$, and $\{0,1\}$ with probability $1-p$. 
A cluster is thus a random subcube of $\{0,1\}^N$. If $\pi^A_i=\{0\}$ or $\{1\}$, variable $i$ is said ``frozen'' in $A$; otherwise it is said ``free'' in $A$.
One given configuration $\vec \sigma$ might belong to zero, one or several clusters. A solution belongs to at least one cluster. 

The parameter $\alpha$ is analogous to the density of constraints in CSPs; clearly the SAT-UNSAT transition occurs at $\alpha_s := 1$, where clusters cease to exist.
The parameter $p$ gives the probability that a variable is frozen, and plays a role similar to the clause size $k$ in $k$-SAT, or to the number of colors $k$ in $k$-coloring, as we will see later.
Note that in the special case $p=1$, the RSM is equivalent to the random code model with rate $R=1-\alpha$ \cite{Montanari01}. 

Frozen variables and the structure of the solution space have been introduced to mimic the situation observed in random CSPs. However there are important differences between the RSM and models like $k$-SAT.
First, in real CSPs the clusters are not necessarily subcubes of $\{0,1\}^N$. We stress here that when speaking about clusters in the RSM we have in mind the above definition, whereas in the context of the CSPs the notion of cluster is more general \cite{MezardParisi02,KrzakalaMontanari06}. Further, in real CSPs the sets of frozen variables associated with clusters are correlated by the underlying graph, instead of being distributed uniformly. Moreover, free variables in CSPs do not enjoy the same freedom as in the RSM, as clusters usually do not fill up the whole subcube allowed by the frozen variables. In fact, free variables can be correlated within each cluster in a highly nontrivial way, and these correlations may even be so strong that they create clusters without the help of frozen variables. Clusters without frozen variables are indeed very important, as discussed recently in \cite{ZdeborovaKrzakala07,Semerjian07}.

\subsection{The basic structural phase transitions}

We now describe the static properties of the RSM in the thermodynamic limit $N\to \infty$ (the two parameters $\alpha$ and $p$ being fixed and independent of $N$). The internal entropy $s$ of a cluster $A$ is defined as $\frac{1}{N}\log_2 |A|$, i.e. the fraction of free variables in $A$. 
The probability ${\cal P}(s)$ that a cluster has internal entropy $s$ 
follows the binomial distribution
\be
  {\cal P}(s) = {N\choose sN} (1-p)^{sN} p^{(1-s)N}\, .     
\ee 
Let ${\cal N}(s)$ be number of clusters of entropy $s$. This number follows a binomial law of parameter ${\cal P}(s)$ with $2^{N(1-\alpha)}$ terms.
Then the mean and the variance of ${\cal N}(s)$ read: 
\be
         \mathbb{E} {\cal N}(s)= 2^{N(1-\alpha)} \cP(s)\, , \quad \quad   
            {\rm Var}{\cal N}(s)= 2^{N(1-\alpha)} \cP(s) [1-\cP(s)]\, . \label{moments}
\ee
By Markov's inequality:
\be
\bP\left[\cN(s)\geq 1\right]\leq \E\left[\cN(s)\right],
\ee
and by Chebyshev's inequality:
\be
      \mathbb{P}\left\{ \left| \frac{{\cal N}(s)}{\mathbb{E} {\cal N}(s)} -1 \right|>\varepsilon \right\} \le \frac{{\rm Var}{\cal N}(s)}{[\mathbb{E} {\cal N}(s)]^2\varepsilon^2 } \leq \frac{1}{2^{N(1-\alpha)}\varepsilon^2  \cP(s)}\quad\textrm{for all }\varepsilon>0, \label{Chebyshev}
\ee
we get, with high probability (w.h.p.: with probability going to $1$ as $N\to\infty$):
\begin{equation}
\lim_{N\to\infty}\frac{1}{N}\log_2\cN(s) =\left\{ \begin{array}{ll}  \Sigma(s):= 1-\alpha-D(s\parallel  1-p ) & \textrm{ if }\Sigma(s)\geq 0,\\
-\infty & \textrm{ otherwise,}\end{array}\right.
\label{sigma}
\end{equation}
where $D(x\parallel  y) := x\log_2{\frac{x}{y}}+(1-x)\log_2{\frac{1-x}{1-y}}$ is the binary Kullback-Leibler divergence. Throughout the paper, the same Markov/Chebyshev argument will apply every time we will have to deal with a number of clusters with a specific property.

We now compute the total entropy $s_{\rm tot}=\frac{1}{N}\log_2 |S|$. First note a random configuration belongs on average to $2^{N(1-\alpha)} (1-\frac{p}{2})^N$ clusters. Therefore, if
\be\label{eq:defalphad}
\alpha< \alpha_d := \log_2{(2-p)},
\ee
then with high probability the total entropy is $s_{\rm tot}=1$.

Now assume $\alpha> \alpha_d$. The total entropy is given by a saddle-point estimation:
\be 
\sum_{A} 2^{s(A)N} = [1+o(1)] N\int_{\Sigma(s)\geq 0} \ud s\ 2^{N[\Sigma(s)+s]},\label{eq:calcstot}
\ee
\be
\textrm{whence}\quad     s_{\rm tot}   =  \max_{s}{\left[\Sigma(s)+s\,|\,\Sigma(s)\ge 0\right]}.
     \label{s_tot}
\ee
We denote by $s^*={\rm argmax}{[\dots]}$ the fraction of free variables in the clusters that dominate the sum. Note that in this sum solutions belonging to several clusters have been counted too many times. This does not affect the validity of our estimation, since in every cluster the fraction of solutions belonging to more than one cluster is exponentially small as long as $\alpha>\alpha_d$.


Define $\tilde s:=2(1-p)/(2-p)$ such that $\partial_s \Sigma(\tilde s)=-1$.
The complexity of clusters with entropy $\tilde s$ reads:
\be
      \Sigma(\tilde s) = \frac{p}{2-p} + \log_2(2-p) - \alpha .
\ee
$\tilde s$ maximizes Eq.~(\ref{s_tot}) as long as $\Sigma(\tilde s)\ge 0$, 
that is if
\be 
\alpha\leq \alpha_c:= \frac{p}{(2-p)}+\log_2{(2-p)}.
\ee
Then the total entropy reads
\be
        s_{\rm tot} = 1 - \alpha + \log_2{(2-p)} \, \quad {\rm for} \quad \alpha \le \alpha_c.
\ee
For $\alpha>\alpha_c$, the maximum in (\ref{s_tot}) is realized by the largest possible cluster entropy $s_M$, which is given by the largest root of $\Sigma(s)$. Then $s_{\rm tot}=s^*=s_M$.
In this phase the dominating clusters\footnote{The ``dominating clusters'' are the minimal set of clusters covering almost all solutions.} have size $e^{Ns^*+\Delta}$, where $\Delta=O(1)$ is asymptotically distributed according to a Poisson point process of rate $e^{-m\Delta}$, i.e., for ${\rm d}\Delta \ll \Delta$ the probability that there is at least one state of size between $e^{Ns^*+\Delta}$ and $e^{Ns^*+\Delta+{\rm d}\Delta}$ is $e^{-m\Delta}{\rm d}\Delta$, where $m=-\partial_s \Sigma(s^*)$. 
Extreme value study of this process leads to the Poisson-Dirichlet \cite{MezardParisi84,Talagrand00,PitmanYor97} distribution of weights of clusters. In particular it follows that an arbitrary large fraction of the solutions can be covered by a finite number of clusters. Such a phase is called condensed.


\begin{figure}
\begin{center}
\resizebox{.99\linewidth}{!}{\input{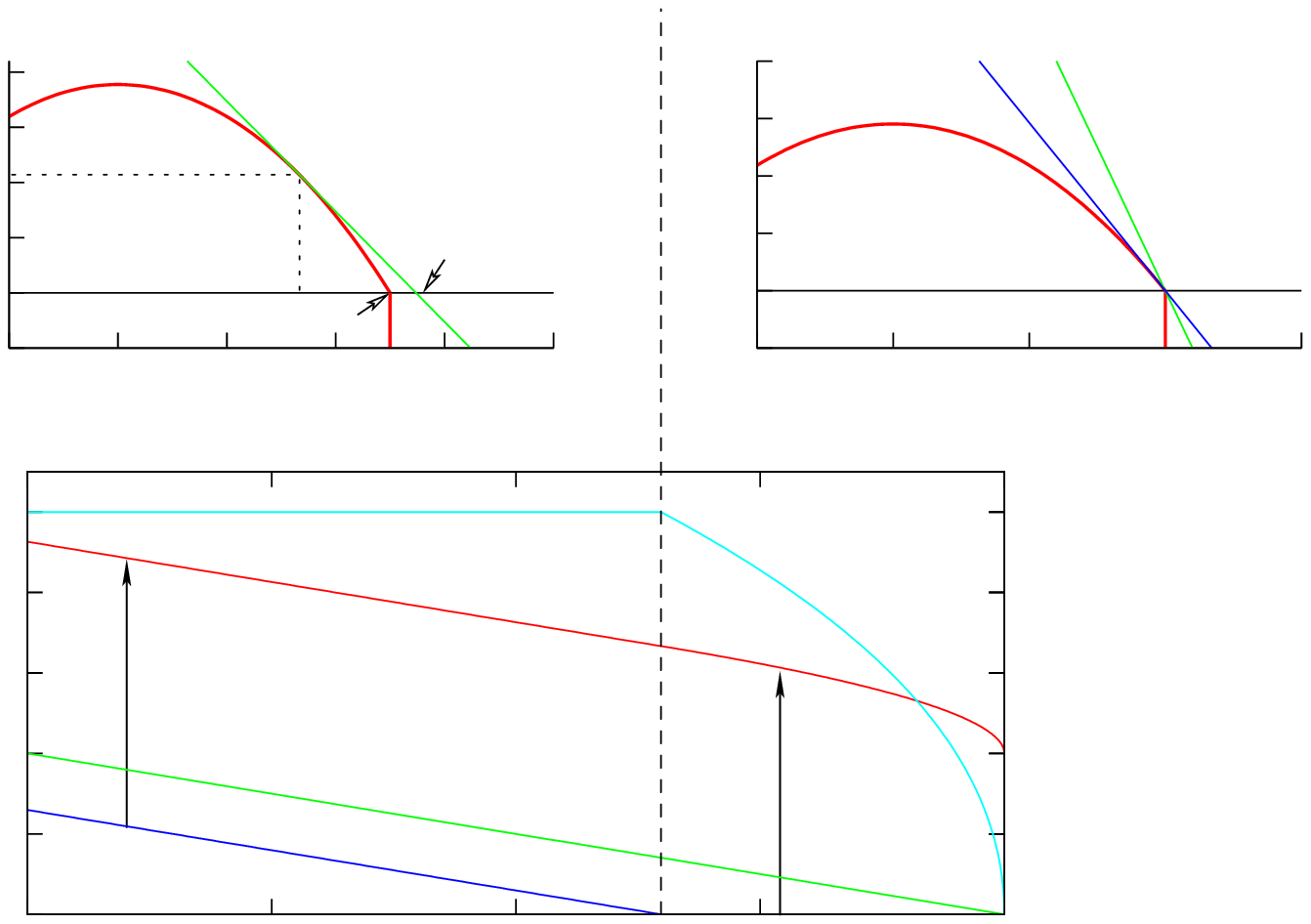}}
\caption{\label{fig:diagphase}(Color online) Top: graphical construction of the maximum of $\Sigma(s)+s$ by a Legendre transformation. In the top left figure ($\alpha<\alpha_c$), the line of slope $-1$ tangent to the complexity function gives the saddle-point $s^{*}$, as well as the total entropy $s_{\rm tot}$ by the intercept on the $s$ axis. In the top right figure ($\alpha>\alpha_c$), the supporting line of slope $-1$ gets ``stuck'' at $s_M$, where the derivative is $-m>-1$. Bottom: represented as a function of $\alpha$: total entropy $s_{\rm tot}$, total complexity $\Sigma_{\rm tot}=1-\alpha$, typical entropy $s^*$, complexity of dominating clusters $\Sigma^*=\Sigma(s^*)$, and $m=-\partial_s \Sigma(s^*)$. The condensation point $\alpha_c$ marks the separation between the two regimes illustrated above.}
\end{center}
\end{figure}

In summary, for a fixed value of the parameter $p$, and for increasing values of $\alpha$, four different phases can be distinguished:
\begin{itemize}
   \item[(a)]{Liquid phase, $\alpha < \alpha_d$: almost all configurations are solutions.}
   \item[(b)]{Clustered phase with many states, $\alpha_d < \alpha < \alpha_c$: 
an exponential number of clusters is needed to cover almost all solutions.}
   \item[(c)]{Condensed clustered phase, $\alpha_c < \alpha < \alpha_s=1$: 
a finite number of the biggest clusters cover almost all solutions.}
   \item[(d)]{Unsatisfiable phase, $\alpha> \alpha_s$: no cluster, hence no solution, exists.}
\end{itemize}
The very same series of phase transitions is observed in random $k$-coloring and random $k$-satisfiability, where $\alpha$ is the density of constraints \cite{KrzakalaMontanari06,ZdeborovaKrzakala07}. The condensation transition at $\alpha_c$ corresponds to the Kauzmann temperature \cite{Kauzmann48} in the theory of glasses; at $\alpha_c$ the total entropy $s_{\rm tot}(\alpha)$ has a discontinuity in its second derivative with respect to $\alpha$, in analogy with the discontinuity of the specific heat at the Kauzmann temperature. 

Keeping in mind the similarity between the RSM and real CSPs, it can be useful to mention some of the properties that are commonly discussed in the statistical physics analysis of these problems (for brevity we omit the proves of these statements). Among them is the probability distribution of mutual overlaps $P(q)$, where $q(\bs,\bs')=\sum_{i=1}^N [2\delta(\sigma_i,\sigma'_i)-1]/N$. Below the condensation transition, $\alpha<\alpha_c$, we have $P(q)=\delta{(q)}$, reflecting the fact that random pairs of solutions are uncorrelated. For $\alpha>\alpha_c$, the overlap function consists of intra-cluster and inter-cluster overlaps: $P(q)= w \delta[q- (1-s_{\rm tot})] + (1-w) \delta(q)$, where $w$ is the sum of squares of weights of all the clusters, it is a non-self-averaging random variable, the distribution of which can be computed from the Poisson-Dirichlet process \cite{PitmanYor97,GrossMezard84}.

An equivalent way of characterizing the condensed phase is to consider the $k$-point correlation function, with $k\geq 2$:
\be 
    \sum_{x_1,\ldots,x_k} | \mathbb{P}(\sigma_1=x_1,\sigma_2=x_2,\dots,\sigma_k=x_k) - \mathbb{P}(\sigma_1=x_1)\dots \mathbb{P}(\sigma_k=x_k) |.
\ee
This quantity decays to zero as $N$ goes to infinity in the non-condensed phase, whereas it remains bounded away from zero in the condensed phase.


\subsection{The large $k$-limit of random $k$-SAT and $k$-COL}
\label{sec:largeK}

One of the most interesting properties of the random-energy model \cite{Derrida80} is its equivalence with the large-$p$ limit of the $p$-spin glass \cite{GrossMezard84}. In the same spirit, although the justification is slightly different, the random-subcube model is found to be equivalent to random $k$-SAT, random $k$-COL, and presumably other constraint satisfaction problems in the limit $k \to \infty$, for connectivities close to the satisfiability threshold. Let us detail this statement.

It was already known that at zeroth order (when $k\to \infty$) random $k$-SAT and $k$-COL behave as a random-code model (random-subcube model with $p=1$), in which clusters are uniformly distributed singletons. Recent large-$k$ calculations of the cluster size distribution $\Sigma(s)$ in $k$-SAT and $k$-COL \cite{KrzakalaMontanari06,ZdeborovaKrzakala07} allow a direct comparison with the RSM.

The control parameters of the RSM are rescaled as: 
\be
      p=1-\varepsilon\, , \quad \alpha = 1+ \varepsilon \frac{1+\gamma}{\ln 2},
\ee
with $\varepsilon \ll 1$ and $\gamma=\Theta(1)$. The cluster size distribution in the RSM is then, at leading order:
\be
\Sigma(s) \ln(2) = s \left[1-\ln{\frac{s}{\varepsilon}}\right] -\varepsilon(2+\gamma) + o(\varepsilon).
 \label{largeK}
\ee
while the condensation and satisfiability thresholds read, in terms of the rescaled variable $\gamma$:
\be
      \gamma_c= -2\ln{2}\, ,  \quad \quad \gamma_s=-1.  \label{gam_c}
\ee

In $k$-SAT and $k$-COL, identifying $\varepsilon$ and $\gamma$ in the following way:
\bea
{\rm SAT:} \quad  \varepsilon&=&\frac{1}{2^{k+1}}\, , \quad \frac{M}{N}=2^k\ln{2}-\frac{\ln{2}}{2} +\frac{\gamma}{2}\, , \\ 
{\rm COL:} \quad  \varepsilon&=&\frac{1}{2k}\, , \quad \frac{M}{N}=k\ln{k}-\frac{\ln{k}}{2} +\frac{\gamma}{2} \, ,
\eea
where $M$ is the number of constraints (edges in coloring), and $N$ the number of variables, gives a perfect match\footnote{Note the difference in the logarithmic base between here and \cite{KrzakalaMontanari06,ZdeborovaKrzakala07}.} for the complexity function \eqref{largeK}, as computed in \cite{KrzakalaMontanari06,ZdeborovaKrzakala07}. This illustrates the analogy between $\alpha$ and the density of constraints $M/N$, as well as between $p$ and $k$.

This equivalence goes further than simply having identical cluster size distributions. The cavity analysis of $k$-SAT shows that the fraction of free variables in a cluster scales exactly as its internal entropy $s$.
The entropy of clusters is thus maximal, from which we infer that clusters fill up the whole subcube prescribed by their frozen variables, like in the RSM. The same is true for $k$-COL (compare Eqs.~(E14) and (E27) in \cite{ZdeborovaKrzakala07}) with the small difference that for every unfrozen variable only two (out of $k$) colors are allowed.
   
Note that this comparison is valid only in a finite vicinity of the satisfiability threshold, for $\gamma=\Theta_k(1)$. In particular, it does not encompass the clustering transition, which for $k$-SAT (resp. $k$-COL) occurs for constraint densities scaling as $2^k \ln 2 /k$ (resp. $k \ln k /2$).

\subsection{Decimation}

An important contribution of statistical physics to the field of combinatorial optimization has been to exploit the information provided by message-passing algorithms to devise physics-guided decimation schemes. 

Message-passing algorithms exchange information between units (variables and constraints) in order to obtain {\em estimates} of marginal probabilities (beliefs) or other related quantities (e.g. surveys, see below). Subsequently, this information is used to find a solution.
A usual way to do this, called decimation, proceeds as follows: fix randomly the value of one\footnote{In practice the number of variables fixed at each step can range from one to a small fraction of the variables.} variable according to its estimated belief, then re-run the message-passing algorithm on the reduced system, and loop. A trivial statement is that a perfect estimate of all marginal probabilities would always cause the decimation procedure to find a solution (if any).

In the RSM there is no underlying graph, therefore message-passing cannot be defined. 
However, it is possible to study decimation schemes based on exact marginal probability estimators. Although such procedures should really be viewed as thought experiments, they can be used to gain some insight on real algorithms.
Here two idealized algorithms are considered:
\begin{itemize}
\item Belief estimator: outputs the exact marginal probabilities $\mu_i(\sigma_i)=\sum_{\bs_{\backslash i}} \mu(\bs)$, where $\mu(\bs)=\mathbb{I}(\bs\in S)/|S|$.
\item Survey estimator: outputs ``surveys'', i.e. marginal probabilities over the clusters: $\nu_i(\pi_i)=\sum_{\bpi_{\backslash i}} \nu(\bpi)$, where $\nu(\bpi)=\sum_A \mathbb{I}(\bpi=\bpi^A)/\lfloor 2^{N(1-\alpha)}\rfloor$.
\end{itemize}

In real CSPs, belief and survey propagation arguably provide {\em asymptotically} accurate estimators, as long as the number of clusters dominating the measure $\mu$ (or $\nu$ for survey propagation) scales exponentially with $N$, and as long as the one-step replica symmetry breaking description is the correct one \cite{MezardParisi01,ZdeborovaKrzakala07}. However, belief-guided and survey-guided decimation schemes are difficult to analyze in real CSPs ---\,see \cite{Parisi03} for an empirical study on surveys and \cite{MontanariRicci07} for recent analytical study on beliefs in $k$-SAT. 

Let us study decimation in the RSM. As long as the phase is non-condensed, the belief estimator always outputs $\mu_i(\sigma_i)\approx 1/2$ for all $i$ in the limit $N\to \infty$. Likewise, the survey estimator will output $\nu_i(\{0\})\approx \nu_i(\{1\})\approx p/2$, $\nu_i(\{0,1\})\approx 1-p$. In both cases, the decimation procedure is completely unbiased: it will fix a random variable $i$ to $0$ or $1$, with probability $1/2$. This observation remains true in the subsequent decimation steps as long as the number of clusters dominating the reduced measure $\mu$ (or $\nu$) remains exponential. Within this assumption, after $T=tN$ ($0 \le t \le 1$) decimation steps, $T$ variables will be fixed randomly and independently. We are then left with a restricted space of solutions compatible with these $T$ fixed variables. The logarithm of the number of clusters of entropy $sN$ is then
\be 
 N\Sigma_t(s) =N\left[ 1-\alpha + t \log_2\left(1-\frac{p}{2}\right) - (1-t) D\left(\frac{s}{1-t}||1-p\right)\right].
\ee  
Rescaling by the number of unfixed variables $(1-t)N$: $\Sigma_t=(1-t)\bar\Sigma_t$, $s=(1-t)\bar s$, we obtain
\be
  \bar \Sigma_t(\bar s) = 1-\frac{\alpha - t \alpha_d}{1-t} -D(\bar s||1-p).
\ee
The parameter $\bar \alpha(t) := (\alpha - t \alpha_d)/(1-t)$ now plays the same role as $\alpha$ in the analysis of the RSM. Consequently, the system undergoes the same condensation and unsatisfiability transitions as $t$ is increased.
Assume for example that $\alpha_d < \alpha <\alpha_c $. Fixing a fraction $t_c:=(\alpha_c-\alpha)/(\alpha_c-\alpha_d)$ of the variables will cause the system to condense. The belief estimator will then be dominated by a finite number of clusters, yielding instance-dependent biases on variables. 
An extensive number of variables become suddenly near-frozen, i.e., $\mu_i(1)\approx 0$ or $\approx 1$, and remain so for $t>t_c$. At $t_s:=(1-\alpha)/(1-\alpha_d)$, the total complexity goes to zero, and near-frozen variables become truly frozen, i.e. $\mu_i(1)=0$ or $1$, as all sub-dominant clusters disappear.
By contrast, the survey estimator will output unbiased marginal probabilities as long as the total number of clusters is exponential in $N$, that is if $t<t_s$. At $t= t_s$, decimation concentrates on a single cluster, causing a freezing avalanche in the surveys, i.e. for each variable $i$, $\nu_i(\{0,1\})=1$ or $\nu_i(\{0\})=1$ or $\nu_i(\{1\})=1$. For both estimators, any slight error at $t_s$ will cause the failure of the decimation process.

Note that in real CSPs, belief propagation is not expected to be correct in the condensed phase, beyond $t_c$; in fact, it will not detect condensation, nor near-frozen variables.
For a recent study of the belief-propagation-guided decimation in $k$-SAT see \cite{MontanariRicci07}.
Remarkably, in real CSPs, survey-guided decimation usually simplifies the problem: after a certain number of decimation steps, it outputs $\nu_i(\{0,1\})=1$ for all $i$, and the problem can then be passed over to a simple local search algorithm \cite{MezardZecchina02,Parisi03}. Although the RSM, which is intrinsically clustered, is unable to capture this property, it sheds some light on why decimation may not work in some cases.




\section{Distance and ergodicity}
\label{two}

The geometrical organization of solutions to CSPs is thought to play an important role in setting intrinsic limits to the performance of search algorithms. In particular, the clustering phenomenon, by which the solution space is fragmented into many connected components far from each other, has been proposed by physicists as a possible explanation for the failure of most known algorithms \cite{BiroliMonasson00,MezardParisi02,MezardZecchina02}. Also, the role of frozen variables was recently discussed in \cite{ZdeborovaKrzakala07}. Conversely, the success of survey propagation \cite{MezardZecchina02} is usually explained by the fact that it explicitly incorporates the existence of clusters. The separability of clusters has been proved in the $k$-SAT problem, in compliance with the predictions of statistical physics \cite{MezardMora05}.
Despite this evidence, the precise relation between geometry and algorithms still lacks a rigorous base. The random-subcube model offers an excellent opportunity to study these questions in a well controlled framework.

\subsection{The dynamical transition}

Let us first argue why the clustering transition at $\alpha_d$, defined by Eq.~\eqref{eq:defalphad}, actually corresponds to what is commonly refered to as the
dynamical (ergodicity breaking) transition in real CSPs \cite{MontanariSemerjian05,MontanariSemerjian06,MontanariSemerjian06b,KrzakalaMontanari06,MezardRicci03,ZdeborovaKrzakala07}. To this end we study a uniform unbiased random walk on the space of solutions: at each step, one is allowed to move from one solution to the other by flipping only one spin. We choose this dynamics for the sake of simplicity, but most of the arguments below hold for more general dynamical rules, like for example the flipping of a sub-extensive number of spins at each step.

We have already pointed out that an arbitrary configuration belongs to
$
2^{(1-\alpha)N} (1-p/2)^N 
$
different clusters w.h.p. if $\alpha<\alpha_d$, and to none if $\alpha>\alpha_d$. Therefore below $\alpha_d$ almost all configurations are solutions, and any reasonable dynamics will explore the entire phase space uniformly. 

On the other hand, when $\alpha>\alpha_d$, solutions become exponentially rare. Let $A$ be a cluster of internal entropy $s$. What is the probability that after $t$ steps, a random walker ends up in another cluster $B$ of internal entropy $s'$? Let $a$ denote the proportion of variables that are free in $A$ and frozen in $B$. The probability distribution of $a$ is given by the total number of partitions of $\{1,\ldots,N\}$ into four categories: frozen in $A$ and $B$, only frozen in $A$, only frozen in $B$, and frozen neither in $A$ nor $B$. This probability reads:
\begin{equation}
q(a)=\frac{1}{\binom{N}{Ns}\binom{N}{Ns'}}\frac{N!}{(Na)![N(s-a)]![N(1-s'-a)]![N(s'-s+a)]!}.
\end{equation}
In order for the walker to reach $B$ from $A$, it has to match perfectly the prescriptions (freezings) of $B$ on these $aN$ variables. If $t=\Theta(N^d)$, the probability of this happening is $\approx t2^{-aN}$.
Additionally, variables that are frozen in both clusters must coincide, so that $A$ and $B$ have a non-empty intersection. For a random choice of $B$, this happens with probability $2^{N(s'+a-1)}$.

Consequently, the probability that the walker wanders in any other cluster of entropy $s'$ after $t$ steps is union-bounded by:
\begin{equation}
\phi(s\to s')\leq 2^{N\Sigma(s')}\sum_{a} q(a)\, t\, 2^{-aN}\,2^{N(s'+a-1)} 
\end{equation}
The maximum of $\frac{1}{N}\log q(a)$ tends to zero when $N\to\infty$, so that:
\begin{equation}
\limsup_{N\to\infty}\frac{1}{N}\log \phi(s\to s')\leq \Sigma(s')+s'-1
\end{equation}
This quantity remains negative for all $s'$, as long as $\alpha>\alpha_d$. Therefore, in this regime, hopping from one cluster to the other is very unlikely in a sub-exponential number of steps, even though clusters are not disjoint. In this sense we say that ergodicity is broken. Remarkably, the space of solutions becomes non-ergodic as soon as it becomes non-trivial, at $\alpha_d$. 

Let us stress the importance of the entropic barriers between clusters in our analysis. In the real CSPs and optimization problems the energetic barriers are usually thought of as more important, and clusters are even sometimes described as separated (by an extensive distance). The importance of entropic barriers and the possibility of non-extensively separated clusters should, however, not be neglected in the studies of richer models. 

\subsection{$x$-satisfiability}

The notion of $x$-satisfiability was first introduced as a tool to study the geometrical structure of the solution space of CSPs \cite{MezardMora05}. An instance of CSP is said $x$-satisfiable if and only if it admits a pair of solutions separated by a Hamming distance $\sim xN$. In other words, $x$-satisfiability gives the distance spectrum of the solution space. This spectrum is estimated using three quantities:
\begin{itemize}
    \item[a)]{$d_1=x_1(\alpha)N$: the maximum distance between two solutions inside one cluster,}
    \item[b)]{$d_2=x_2(\alpha)N$: the minimum distance between two solutions from two distinct clusters,}
    \item[c)]{$d_3=x_3(\alpha)N$: the maximum distance between any two solutions (presumably from two different clusters).}
\end{itemize}
The first of these quantities is estimated by noting that the maximum distance between any two solutions in a given cluster, i.e. its diameter, equals its entropy $s$. Therefore the maximum diameter/entropy $x_1$ is given w.h.p. by the largest internal entropy $s_M$, i.e. the largest root of $\Sigma(s)=1-\alpha-D(s\parallel 1-p)$.

Now take two clusters $A$ and $B$ at random, and consider the probability that their distance be $xN$. This distance is given by the number of variables which are frozen in both clusters, but in a contradictory way, such that $\pi_A(i)\neq \pi_B(i)$. This happens independently with probability $p^2/2$ for each variable, so that the number of such variables follows a binomial law of parameter $p^2/2$. Therefore, the number $\mathcal{N}(x)$ of pairs of clusters at distance $xN$ coincides w.h.p. with its mean value:
\begin{equation}
\E[\mathcal{N}(x)]=2^{2(1-\alpha)N}\binom{N}{Nx}{\left(1-\frac{p^2}{2}\right)}^{(1-x)N}{\left(\frac{p^2}{2}\right)}^{xN}\asymp 2^{Ns_2(x)}
\end{equation}
if $s_2(x):= 2(1-\alpha)-D(x\parallel p^2/2)>0$, and $\mathcal{N}(x)=0$ w.h.p. if $s_2(x)<0$. Consequently the smallest possible distance between any two clusters is given by $x_2N$, where $x_2(\alpha)$ is the smallest root of $s_2(x)$. A similar argument gives the largest distance between any two solutions from two different clusters: $x_3(\alpha)=1-x_2(\alpha)$.

To sum up, we find that a random instance is $x$-satisfiable w.h.p. if $\alpha<\alpha_s(x)$, and is $x$-unsatisfiable w.h.p. if $\alpha>\alpha_s(x)$, with:
\begin{equation}
\alpha_s(x)=\left\{\begin{array}{ll}
1 & \textrm{if }x\in[0,1-p]\cup [p^2/2,1-p^2/2]\\
1-D(x\parallel 1-p) & \textrm{if }x\in[1-p,x_0]\\
1-\frac{1}{2}D(x\parallel p^2/2) & \textrm{if }x\in [x_0,p^2/2]\\
1-\frac{1}{2}D(1-x\parallel p^2/2) & \textrm{if }x \in [1-p^2/2,1]
\end{array}\right.
\end{equation}
where $x_0$ is solution to $D(x\parallel p^2/2)=2D(x\parallel 1-p)$.

\begin{figure}
\begin{center}
\input{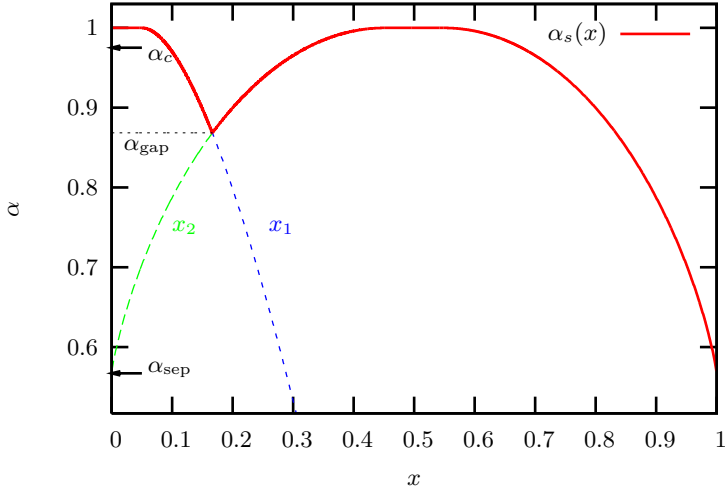}
\caption{\label{fig:xsat}(Color online) The $x$-satisfiability threshold is constructed from the three distances $x_1(\alpha)$, $x_2(\alpha)$ and $x_3(\alpha)$. Below a threshold $\alpha_{\rm gap}$, distance spectra fail to detect the fragmentation of the solution space (there is no more ``gap'' between intra and inter-cluster distances). Below another threshold $\alpha_{\rm sep}$, clusters cease to be all well separated, although ergodicity is still broken. The condensation threshold $\alpha_c$ is shown for information. The dynamical threshold $\alpha_d$ lies outside the picture, and its value is $\approx 0.070$. In this figure $p=0.95$.}
\end{center}
\end{figure}

Fig.~\ref{fig:xsat} shows how $\alpha_s(x)$ can be constructed from the three distances $x_1$, $x_2$, $x_3$. We put $\alpha_{\rm sep}:= 1+(1/2)\log(1-p^2/2)>\alpha_d$, the threshold below which some pairs of clusters have a non-empty intersection. An interesting observation is that ergodicity can still be broken even below this threshold. We can also define the $\alpha_{\rm gap}:=\alpha_s(x_0)>\alpha_{\rm sep}$, below which distances from the same cluster and distances from distinct clusters overlap. This threshold sets the limit below which the notion of $x$-satisfiability fails to detect clustering. The random-subcube model allows us to make a clear and intelligible distinction between the three thresholds $\alpha_d<\alpha_{\rm sep}<\alpha_{\rm gap}$. We expect this distinction to hold in most CSPs\footnote{With the notable exception of $k$-XORSAT, where $\alpha_d=\alpha_{\rm sep}$. Incidentally in $k$-XORSAT we also have $\alpha_c=\alpha_s$.}.

\section{Random energy landscape}
\label{three}

\subsection{Definition}
The random-subcube model can be enriched by adding the notion of energy to the definition of states. The motivation for doing this is to mimic the optimization version of CSPs (where energy is defined as the number of unsatisfied constraints), but it can also be used to reproduce some properties of glassy systems.

For each energy level $E_0$, we define $\mathcal{N}(E_0)=\lfloor 2^{N\Sigma(E_0/N)}\rfloor$ valleys of energy $E_0$, where $\Sigma(e_0)$ is an increasing complexity function. This function can be arbitrary, but for simplicity we will restrict our examples to the form\footnote{This form corresponds to the function used to fit data from the cavity method in the $k$-SAT problem \cite{MezardZecchina02}.} $\Sigma(e_0)=a+be_0-ce_0\ln(e_0)$, where $a>0$ corresponds to a SAT phase, and $a<0$ to an UNSAT phase.

Each valley is defined as a subcube $V$, chosen at random in the same way as clusters in the previous sections, \cf \eqref{eq:defcluster}. In the following, the freezing probability $p$ will be fixed for all energies, but one could easily generalize the model by making it energy-dependent: $p=p(e_0)$.

The number $\cN(E_0,S_0)$ of valleys of energy $E_0=e_0N$ and entropy $S_0=s_0N$ is w.h.p.:
\bea
\begin{array}{lll}
{\cal N}(E_0,S_0)&\asymp  2^{N\Sigma(e_0,s_0)}
&\textrm{if }\Sigma(e_0,s_0):= \Sigma(e_0)-D(s_0\parallel 1-p)\geq 0\\
&=0 &\textrm{otherwise.}
\end{array}
\eea

Given a configuration $\vec \sigma$, we define its energy as a trade-off between the energy of surrounding valleys and their distance. Let us denote the energy of a valley $V$ by $E_0(V)$. Then the energy of $\bs$ is:
\bea
E(\bs):= \min_{V} \left[ E_0(V)+ d(\bs,V) \right]
\label{eq:def_energy}
\eea
where $d(\bs,V)$ is the distance between $\bs$ and the nearest element of $V$. By definition, we say that $\bs$ belongs to the basin of attraction of $V$ if $V$ minimizes the sum.
Observe that with this definition, it may happen that some valleys are not represented at all in the energy landscape. In the following, the term ``state-bottom energy'' shall refer to the energy $E_0$ of the valley minimizing the sum, and the term ``state-bottom entropy'' to the entropy $S_0$ of that valley.


\subsection{Static description of the energy landscape}
What is the energy of an arbitrary configuration $\bs$? Let us start with the typical case: for each $e_0=E_0/N$, we compute the distance to the nearest valley with energy $E_0$. Standard arguments show that the number of valleys of state-bottom energy $e_0N$ at distance $d=\omega N$ is governed by the exponent:
$$
\Sigma(e_0)-D(\omega \parallel p/2).
$$
Then, the minimum distance is given, in the $N\to\infty$ limit, by $\delta[\Sigma(e_0),p/2]N$, where $\delta(x,y)$ is solution to:
$
x=D(\delta\parallel y)
$.
Then, the typical energy is obtained as the best compromise between state-bottom energy and distance:
\bea
e^*=E/N=\min_{e_0}\left\{ e_0+\delta[\Sigma(e_0),p/2]\right\},
\eea
The $\mathrm{argmin}$ gives the typical state-bottom energy $e^*_0$ of a random $\bs$.


As we just saw, most configurations have roughly the same energy, and belong to valleys with the same state-bottom energy. At finite temperature however, thermodynamics will be dominated by configurations of lower energy than $e^*N$. We thus need to estimate the entropy function (governing the number of configurations of given energy), the Legendre transform of which shall give us the free energy.
Given a valley $V$ of energy $E_0=e_0N$ and entropy $S_0=s_0N$, the number of configurations of energy $E$ belonging to this valley is:
\begin{eqnarray}
\mathcal{N}_V(E)&=&2^{S_0}\binom{E-E_0}{N-S_0}\, , \label{eq:s_micro}\\
&\asymp&2^{Ns_V(e|e_0,s_0)},\quad\textrm{with}\quad s_V(e|e_0,s_0):= s_0+(1-s_0)H\left(\frac{e-e_0}{1-s_0}\right)\, ,\nonumber
\end{eqnarray}
where $H(x)=-x\log_2{x}-(1-x)\log_2{(1-x)}$ is the entropy function.
Summing up over all valleys, the total number of configurations with energy $E=eN$ is:
\begin{eqnarray}
\mathcal{N}(E)&=&\sum_{S_0,E_0}2^{S_0}\binom{E-E_0}{N-S_0}2^{N\Sigma(E_0/N,S_0/N)}\asymp 2^{Ns(e)},\\
\quad\textrm{with}\quad s(e)&=&\max_{\substack{e_0,s_0\\ \Sigma(e_0,s_0)\geq 0}}\left[s_V(e|e_0,s_0)+\Sigma(e_0,s_0)\right].\label{eq:sofe}
\end{eqnarray}
Here we have implicitly assumed that all elements in the sphere of radius $E-E_0$ and center $V$ are in the basin of attraction of $V$, as long as $E<e^*N$. This is not true in general, as some configurations may in fact belong to a more favorable basin, and may thus have lower energies. However, such configurations remain exponentially rare in comparison to the total weight of the sphere. Therefore the previous estimate holds\footnote{This argument is similar to the one used in Eq.~\eqref{eq:calcstot}, where some solutions were counted several times, but with no consequence at the exponential scale.}.

Canceling the derivative w.r.t. $s_0$ in \eqref{eq:sofe} yields the saddle for $s_0$:
\bea
\tilde s_0=\frac{(1-p)(1-e+e_0)}{1-p/2} \label{eq:s_bottom}
\eea
Provided that the maximum is reached in a region where $\Sigma(e_0,\tilde s_0)>0$, we get:
\begin{equation}
s(e)=\max_{e_0}\left[1-D(e-e_0\parallel p/2) + \Sigma(e_0)\right]   \quad   {\rm for} \quad e_c<e<e^*.
\end{equation}
This will be valid from $e=e^*$ (for which we find $s(e^*)=1$ as expected) down to a certain condensation energy $e_c$.
Below that energy ($e\leq e_c$), the phase is condensed: $\Sigma(e_0,\tilde s_0)<0$, and the maximum in \eqref{eq:sofe} is reached on the border on the definition domain, where $\Sigma(e_0,s_0)=0$. If we denote by $s_M(e_0)$ the biggest valley of energy $e_0$ (i.e. the largest of root of $\Sigma(e_0,s_0)=0$), we get:
\begin{equation}
s(e)=\max_{e_0}\left\{s_V\left[e|e_0,s_M(e_0)\right]\right\}   \quad   {\rm for} \quad e<e_c.\label{eq:condens}
\end{equation}
So far we have worked in the microcanonical ensemble, but the same arguments hold in the canonical ensemble. In particular the condensation temperature is $T_c={\left(\partial s/\partial e|_{e=e_c}\right)}^{-1}$.
The number of dominating states in the condensed phase follows again a Poisson-Dirichlet process \cite{Talagrand00,PitmanYor97} with parameter $m$, where $m/T$ is the slope of the curve $\Sigma_T(f)$ at its smallest root. The function $\Sigma_T(f)$ is the canonical counterpart of $\Sigma(e_0,s_0)$, and reads: 
\be
      \Sigma_T(f)=\max_{e_0,s_0:f_V(T|e_0,s_0)=f}\Sigma(e_0,s_0)\, ,
\ee
where $f_V(T|e_0,s_0)$ is the single-state free energy, obtained as the Legendre transform of $s_V(e|e_0,s_0)$ in (\ref{eq:s_micro}).

\begin{figure}
\begin{center}
\resizebox{.75\linewidth}{!}{\input{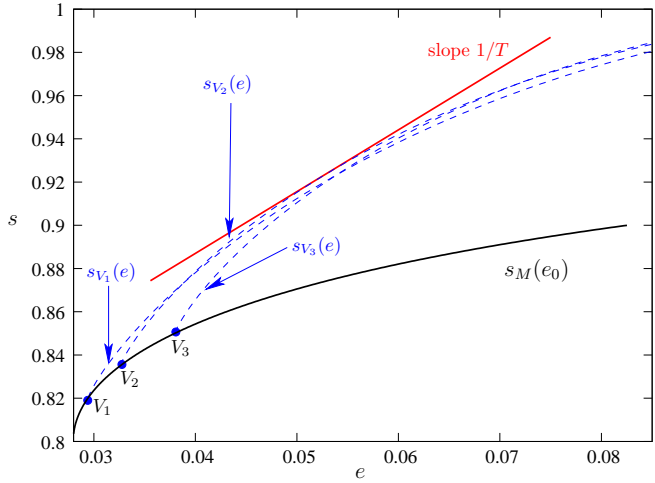}}
\caption{\label{fig:convhull}(Color online) Illustration of the temperature chaos by construction of the microcanonical entropy in the condensed phase, \cf Eq.~\eqref{eq:condens}, and level crossing. We have represented three entropy curves $s_V(e)$ corresponding to three extremal states $V_1$, $V_2$ and $V_3$, the envelope of all these curves is the microcanonical entropy. These states are maximally atypical; they realize a balance between low state-bottom energies and a high state-bottom entropies, which are related by $s_0=s_M(e_0)$ (thick curve). As the temperature (or energy) is decreased, the curves $s_V(e)$ cross each other, and the system is dominated by states of lower state-bottom energies and entropies.  These data were obtained for number of valleys $\Sigma(e_0)= -0.05 - 0.5 e_0 \ln{e_0}$ and $p=0.2$.}
\end{center}
\end{figure}

In both low-temperature phases (condensed and non-condensed), equilibrium is reached for different values of $(e_0,s_0)$ as $e$ varies. Said differently, the states dominating the microcanonical measure at $e$ and at $e+\delta e$ are completely distinct. In the canonical language, we say that the system exhibits temperature chaos \cite{BrayMoore87,KrzakalaMartin02}: slightly changing the temperature from $T$ to $T+\delta T$ dramatically modifies the free-energy landscape, reshuffling the ordering of states. Consequently, correlations are nonexistent between $T$ and $T+\delta T$. This phenomenon of free-energy crossings is illustrated (in the condensed phase) by Fig.~\ref{fig:convhull}, where the maximum of Eq.~\eqref{eq:condens} is constructed geometrically.

\subsection{Relation with dynamics}

We now undertake to describe the dynamical properties of this energetic landscape. To that end we shall make use of the static picture, which we know precisely from bottom-up construction.

Our reasoning proceeds in two steps. First, we study the behavior of a single spin-flip Monte-Carlo dynamics with detailed balance evolving in a {\em single} valley of state-bottom energy $e^*_0$, and state-bottom entropy $s^*_0$. In a second step, we argue that the  same dynamics run on the full rugged energy landscape is entirely governed by this single-state-like behavior.

The thermodynamics of a single typical state is given by:
\begin{equation}
s(e|e_0^*,s_0^*)=s_0^*+(1-s_0^*)H\left(\frac{e-e^*_0}{1-s_0^*}\right)
\end{equation}
where $s_0^*=(1-p)(1-e^*+e^*_0)/(1-p/2)$ is the typical state-bottom entropy of a random configuration, computed from Eq.~(\ref{eq:s_bottom}). Supposing now that the energy landscape is made of this state only, we can easily convince ourselves that the dynamics is ergodic: the energy landscape is convex, and there are neither energetic nor entropic barriers. Both the quenched and annealed\footnote{By quench, resp. annealing, we mean a fast, resp. slow, change of temperature from $T_1>T_d$ to $T_2<T_d$ and some  time (a finite number of Monte Carlo sweeps) spent at $T_2$ afterwards.} dynamics in this simple landscape is then given by the thermodynamics.

We know from the previous paragraphs that states extend up to energy $e^*$, where a ``crest'' connects the different valleys. This crest is actually more of a plateau, as it embeds almost all configurations: $s(e^*)=1$. Therefore, our dynamics will remain there as long as the temperature does not allow configurations of lower energy. This happens at $T_d:=\left(\partial s/\partial e|_{e=e^*}\right)^{-1}$, where exploring valleys starts to be more favorable.

Below $T_d$, the system will find itself trapped in one state, since barriers between valleys are extensive. This randomly picked state has typical properties: in particular, its bottom-energy is $e_0^*$ and its bottom-entropy is $s_0^*$.

We argue that below $T_d$ everything happens as if this trapping state was put in isolation, as we have described above. The justification comes from the fact that a valley does not ``see'' its neighbors as long as $e<e^*$: even though many configurations of energy $e<e^*$ in the isolated state actually belong other valleys in the full landscape, their proportion is exponentially small. In other words, although there may be some directions for which the energy barrier is lower than $e^*-e$, these directions are beaten entropically.

With this reasoning, both the quenched and annealed dynamics are described by the microcanonical entropy of a single typical state: $s_{\rm dyn}(e)=s(e|e_0^*,s_0^*)$.
We note in passing:
\begin{equation}
T_d^{-1}=\left.\frac{\partial s}{\partial e}\right|_{e=e^*}=\left.\frac{\partial s_{\rm dyn}}{\partial e}\right|_{e=e^*},
\end{equation}
as partial derivatives w.r.t. $e_0$ and $s$ inside \eqref{eq:sofe} cancel at the maximum. The dynamical temperature $T_d$ is thus well defined.

Note that this analysis is exact: the fact that it purely relies on static arguments should be attributed to the simplicity of the model.
In richer mean-field models this kind of arguments might also apply, but will most probably not give the full picture.

\begin{figure}
\begin{center}
\resizebox{.8\linewidth}{!}{\input{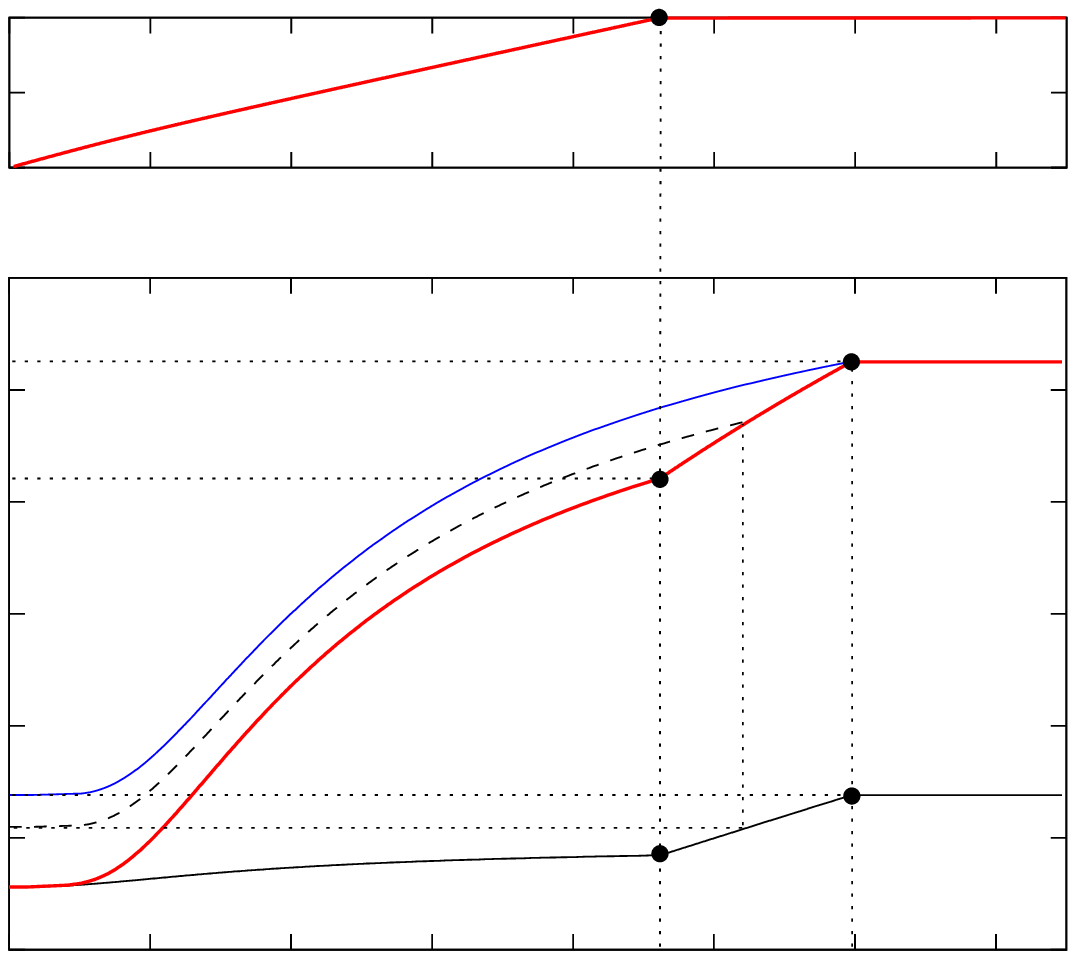}}
\caption{\label{fig:diagTE}(Color online) Energy as a function of temperature. At temperature $T>T_d$, the system is in a liquid state: the dynamics is exploring ergodically all configurations of energy $e^*$. Below the dynamical temperature ergodicity is broken. The upper ``dynamical'' curve shows the result of a quench/annealing in temperature, whereby the systems remains stuck in a typical state of bottom energy $e_0^*$. For $T_c<T<T_d$, equilibrium thermodynamics is dominated by  an exponential number of states (curve ``glass''). Below the condensation (Kauzmann) temperature thermodynamics is dominated by a finite number of states (curve ``condensed glass''). This number is given by a Poisson-Dirichlet process of parameter $m$ (plotted in the upper part of the diagram). The dashed line shows the result of a quench/annealing starting from an equilibrium state at temperature $T_c<T<T_d$. The bottom line shows the bottom energies of the thermodynamically dominating states. These curves were obtained for $\Sigma(e_0)= -0.05 -0.5 e_0 \ln{e_0}$ and $p=0.6$.}
\end{center}
\end{figure}

\subsection{Glassy behaviour}

The analysis of the energy landscape in the RSM is summarized in the energy-temperature phase diagram of Fig. \ref{fig:diagTE}. As anticipated, the behaviour of the RSM resembles the one observed in glasses and spin-glasses. The two distinct glassy transitions (dynamical  and condensation), as well as the phenomenon whereby the physical dynamics gets stuck in metastable states, have been described for example in the $p$-spin glass \cite{GrossMezard84}, the spherical $p$-spin glass \cite{CugliandoloKurchan93}, the Potts glass \cite{GrossKanter85} and the lattice glass \cite{BiroliMezard02}. Several related examples of energy-temperature diagrams were derived recently in \cite{KrzakalaKurchan07}. 

In the aforementioned mean-field models, the static behaviour is better understood than the dynamics. Static properties are usually analyzed by the replica/cavity method, with the help of Parisi's replica symmetry breaking scheme. A satisfactory analytic treatment of the dynamics exists only for the spherical $p$-spin glass \cite{CugliandoloKurchan93}, in which all states have the same entropy. A remarkable step towards connecting the static picture and dynamical behaviour in a rather general framework was done in \cite{MontanariSemerjian05,MontanariSemerjian06,MontanariSemerjian06b}. The RSM could provide a tractable playground for studying several aspects of glassy dynamics, e.g. aging and rejuvenation \cite{LesHouches02}.  

We would like to emphasize the freedom we can enjoy in the definition of the energy landscape. First, arbitrary numbers and sizes of valleys at each energy $e_0$, $\Sigma(e_0)$ and $p(e_0)$, can be considered. Second, the definition of the configurational energy $E(\bs)$ in Eq.~(\ref{eq:def_energy}) could be generalized to an arbitrary function of all valleys $V$ and $\bs$; $E(\bs)={\cal F}(\{V\},\bs)$. By tuning these parameters, one could hope to reproduce the dynamics of more complex models on a quantitative level.

\section{Conclusions}
\label{conclusions}

The random-subcube model is a simple exactly solvable model capturing several interesting properties of random constraint satisfaction problems.
Rather than an attempt to construct a new realistic model for practical instances of constraint satisfaction problems, it allows us to identify which properties of random CSPs can be reproduced by a simple probabilistic structure, and conversely, which of these properties may be intrinsically non-trivial. Examples of reproducible properties include condensation, non-monotony of the $x$-satisfiability threshold, temperature chaos and dynamical freezing in metastable states. From this point of view, the RSM stands just next to the random-energy model \cite{Derrida80}, the random-code model \cite{Shannon48,Montanari01,BargForney02} or the random-energy random-entropy model \cite{KrzakalaMartin02}. 

Since the relation between the RSM and the large-$k$ limit of random $k$-SAT and $k$-COL is based on non-rigorous results from \cite{KrzakalaMontanari06,ZdeborovaKrzakala07}, it would be interesting to establish this equivalence rigorously. Further, the RSM should be helpful for understanding some properties which are too difficult to study in more realistic models, such as finite-size corrections or some aspects of glassy dynamics.

Finally, our work addresses the broad question of producing, inferring, and representing complex and rugged structures of the hypercube.
Although we retained the simplest choice of subcubes for clusters, more sophisticated alternatives could be explored and used to reproduce detailed geometrical features of solution spaces in CSPs. 


\section{Acknowledgment}

We would like to thank Florent Krz\k{a}ka{\l}a and Marc M\'ezard for fruitful discussions about this work, and Guilhem Semerjian and Ji\v r\'{i} \v Cern\'y for critical reading of the manuscript. This work has been partially supported by EVERGROW (EU consortium FP6 IST).

\bibliographystyle{unsrt}
\bibliography{myentries}

\end{document}